\documentclass[nofigure]{pasj00}
 \tenpoint

\SetRunningHead{J.H. Fan et al.}{Beaming Effect in Fermi Blazars}
\title{Beaming Effect in Fermi Blazars}
\author{J. H. Fan$^{1,2}${\thanks{email:jhfan$_{-}\rm{cn}$@yahoo.com.cn}}, J.H. Yang$^{3,1}$, J.Y. Zhang$^{1,2}$, T.X. Hua$^1$,  Y.
Liu$^{1,2}$, Y. P. Qin$^{1,2}$, Y. Huang$^{1,2}$}
\affil{1. Center for Astrophysics, Guangzhou University, Guangzhou 510006, China \\
2. Astronomy Science and Technology Research Laboratory of\\
Department of Education of Guangdong Province, Guangzhou 510006,
China\\
3. Department of Physics and Electronics Science, Hunan University
of Arts and Science, Changde 415000, China}

\Received{$\langle$reception date$\rangle$}
\Accepted{$\langle$acception date$\rangle$}
\Published{$\langle$publication date$\rangle$}
\begin{document}

 \maketitle

\begin{abstract}

The $\gamma$-ray loud  blazars (flat spectrum radio quasars--FSRQs
and BL Lacertae objects-BLs) are very bright in the $\gamma$-ray
bands, which is perhaps associated with a beaming effect.
Therefore, one can expect that the $\gamma$-ray luminosity is
correlated with the beaming factor.
 In this paper, we investigated  the relation between the radio Doppler factors and
the gamma-ray  luminosities.  Our analysis suggests that the  $\gamma$-ray
luminosity be strongly correlated with the factor of $\delta_R$ for the whole sample,
FSRQs, and BLs.   When the effect of a common redshift is excluded,
  the correlation still exists  for the FSRQs sub-sample suggesting
   that the $\gamma$-rays are strongly beamed.
 However, the partial correlation analysis does not show a
      correlation for the small BL Lac sample.

\end{abstract}

\section{Introduction}

Active galactic nuclei (AGNs) are very interesting, their
specially  observational properties have attracted many
astronomers. Blazars are an even extreme subclass of AGNs. There
are two subclasses for blazars: flat spectrum radio quasars(FSRQs)
and BL Lacertae objects (BLs),  the latter  can also be classified
further as radio selected BL Lacertae objects (RBLs) and X-ray
selected BL Lacertae objects-XBLs from surveys, or high-peaked BL
Lacertae objects (HBLs) and
 low-peaked BL Lacertae   objects (LBLs) from spectral energy distribution-SED.
  Blazars show rapid and large
variability, high and variable polarization, superluminal motions
in their radio components, and strong $\gamma$-ray emissions, etc.
(e.g.
 Abdo et al. 2009, 2010a;
 Aller et al.  2011;
 Bastieri 2011;
 Cellone et al. 2007;
 Ciprini et al. 2007;
 Fan et al. 1996, 2011;
 Fan 2012;
 Ghisellini et al. 2010;
 Gupta 2011;
 Gupta et al. 2004;
 Marscher et al. 2011;
 Romero et al. 2002;
 Wills et al. 1992;
 Wagner 2010;
 Urry 2011;
 Yefimov  2011).

During the EGRET mission, about 60 strong $\gamma$-ray emitters
were detected with high confidence (Hartman et al. 1999). However,
the new generation of $\gamma$-ray mission, the Fermi detected a
lot of blazars (see Abdo et al 2010a, Ackermann, et al.  2011a).
Many interesting results have been come to light although the
highly energetic emissions are not very clear (
 Abdo et al. 2010b;
 B\"ottcher  et al. 2008;
 Dermer et al. 2009;
 Ghisellini et al. 2009;
 Graff  et al. 2008;
 Hovatta et al. 2009; and
 Lott 2010).
The bright $\gamma$-ray emissions and the detected variability
suggest that the $\gamma$-rays are strongly beamed.
 Dondi \& Ghisellini 1995,
 Muecke et al. (1997),
 Fan et al. (1998),
 Huang, et al. (1999),
 Cheng et al. (2000), and
 Pushkarev et al. (2010) investigated the correlation between the $\gamma$-ray and the radio
 bands. The correlation suggests an indirectly  beaming effect in the $\gamma$-ray
 emissions.
 Arshakian et al. (2010) investigated the correlation between
the $\gamma$-ray luminosity and the rest-frame radio loudness,
$\rm R = S_{VLBA}/S_{opt}$ (where $\rm S_{VLBA}$ is the VLBA flux
density at 15 GHz and $\rm S_{opt}$ is the optical flux density at
  5100\AA) for some $\gamma$-ray loud blazars and found a significant positive
correlation, which suggests  that the strong $\gamma$-ray jets
have progressively  high Doppler factors (or faster speeds) in the
radio domain compared to those in the optical regime.
  Kovalev et al. (2009) found that  the median brightness temperature T$_b$ values for
 Fermi-detected sources are statistically higher than those for the
 rest of their sample at a 99.9 percent confidence.
 Savolainen et al. (2010) considered 62 objects with apparent velocity
 from MOJAVE and Doppler factors  from radio variability from Metsahovi Radio Observatory,
 and compared the sources detected by Fermi and those not-detected. They
 found that the Fermi-detected blazars have on average higher Doppler factors
 than the non-Fermi-detected blazars.
 We  found that the $\gamma$-ray luminosity is associated with the core-dominance parameter(Fan et al. 2010),
 the $\gamma$-ray variability index is correlated with that in the radio band (Fan et al. 2002), and
 the $\gamma$-ray Doppler factor can be estimated from the radio bands (Zhang, Fan, \& Cheng,
 2002). All those suggest the beaming effect in the $\gamma$-ray
 emissions.

 As proposed by Dermer (1995), the
dependence of the $\gamma$-ray flux on the Doppler factor can be used to
investigate
  the $\gamma$-ray emission mechanism,
different emission mechanism has different dependence of the flux
density ($\rm{S_{\gamma}}$) on the $\gamma$-ray Doppler factor
($\delta_{\gamma}$), namely $\rm S_{\gamma} \propto
\delta^{3+\alpha}$ for
 a synchrotron self-Compton (SSC) model,
  and $\rm S_{\gamma} \propto
\delta^{4+2\alpha}$ for
 an external Compton (EC)
 model.
 These indexes (3+$\alpha$ and 4+2$\alpha$) are true for
 transient emission features, whereas in a steady jet, the indexes are smaller by one, namely,
 2+$\alpha$ and 3+2$\alpha$.
   Therefore, one can use that
dependence  to discuss statistically the $\gamma$-ray emission
mechanism. Unfortunately, the $\gamma$-ray Doppler factors are not
available for any $\gamma$-ray sample.

  If the Doppler factor in the $\gamma$-ray
region, $\delta_{\gamma}$, is the same as that in the radio band,
$\delta_R$, then the $\delta_R$ can be used to deal with the
beaming effect in the $\gamma$-ray
  region and to investigate $\gamma$-ray emission mechanisms.
  Actually, the radio Doppler
factors are not easy to estimate although many methods have been
proposed (see L\"{a}hteenm\"{a}ki \& Valtaoja 1999 for a
comparison). L\"{a}hteenm\"{a}ki \& Valtaoja (1999) proposed to
decompose each flux curve into exponential flares, calculated the
variability time scale of the flare and the corresponding
brightness temperature, $T_{B, var}$. The variability radio
Doppler factor can be estimated using
$\delta_{var}=(\frac{T_{B,var}}{T_{B,in}})^{1/3}$, here $T_{B,in}
= 5\times10^{10}K$ is adopted (see Readhead 1994,
L\"{a}hteenm\"{a}ki et al. 1999).

  In the present paper, we used the available radio Doppler
 factor, $\delta_{\rm R}$,  and the $\gamma$-ray luminosity
 calculated from the data given in the paper (Abdo et al. 2010a)
  to investigate the dependence of the $\gamma$-ray luminosity on
 the radio Doppler factor.
    This paper is arranged as follows: in the 2nd section, we show the
sample and the results; in the 3rd section, we will give some
discussions and a  brief conclusion. We adopt H$_0 = 73 \rm{km
\cdot s^{-1}  Mpc^{-1}}$,  and the spectral index, $\alpha$ is
defined as $f_{\nu} \propto \nu^{-\alpha}$ through this paper.

\section{Sample and Results}

 From Dermer (1995), for the $\gamma$-ray flux density, we
have that
 $\rm S_{\gamma} \propto \delta^{3+\alpha}$ for an SSC model, and
 $\rm S_{\gamma} \propto \delta^{4+2\alpha}$ for an EC model.
  These indexes (3+$\alpha$ and 4+2$\alpha$) are true for
 transient emission features, whereas in a steady jet, the indexes are smaller by one,
 namely,  2+$\alpha$ and 3+2$\alpha$. In the present work, we will only consider  the former case.
   If  the $\gamma$-ray luminosity is taken into account, then we should expect that\\
 $$\rm L_{\gamma}  \propto \delta^{4+\alpha} \,\, for\,\, the\,\, SSC \,\,model,\,\, and $$
 $$\rm L_{\gamma} \propto \delta^{5+2\alpha}\,\, for\,\, the\,\, EC\,\,
 model.$$
In the following sections, we will compile a sample with available
radio Doppler factors and the $\gamma$-ray detections, and then
discuss the dependence of the $\gamma$-ray luminosity on the
Doppler factors.

\subsection{Sample}

Based on the catalogue of 1FGL (Abdo et al. 2010a), we compiled
the available radio Doppler factors  from three literatures,
namely,
    L99: L\"{a}hteenm\"{a}ki \& Valtaoja (1999);
    H09: Hovatta et al. (2009); and
    F09: Fan et al. (2009).  Those radio Doppler factors were determined by the same
    method, but Doppler factors in Fan et al. (2009) are
  based on 8 and 15 GHz radio flux monitoring by Uni. of Michigan
  Radio Observatory whereas those in L\"ahteenm\"aki \& Valtaoja
  (1999) and Hovatta et al. (2009) are based on 22 and 37 GHz
  observations by Metsahovi Radio Observatory.
     There are 59 sources, the
     corresponding data are listed in Table 1. In the Table,
 column (1) gives the name of the source,
 column (2) classification,  H for HBL, L for LBL, F for FSRQ,
 column (3) the redshift,
 column (4) the $\gamma$-ray  photon flux
       in 1-100 GeV in units of photon/cm$^2$/s from Abdo et al. (2010a),
 column (5) the photon spectrum index from Abdo et al. (2010a),
 column (6) the $\gamma$-ray luminosity  in erg/s,
 column (7) radio Doppler factor, $\delta_R$ from L99,
 column (8) radio Doppler factor, $\delta_R$ from H09, and
 column (9) radio Doppler factor, $\delta_R$ from F09.

\begin{table*}
\begin{center}
 Table~ 1. \hspace{20pt} A Sample of 59 Fermi Blazars with Radio Doppler Factors
  \vspace{3pt}
\begin{tabular}{lcccccccc}
\hline
 Name & Class & z & F(1-100GeV)&
$\alpha_{\gamma}$ &log L$_{\gamma}$ &
  $\delta_R^{L99}$ &
  $\delta_R^{H09}$ &
  $\delta_R^{F09}$
 \\
    (1)  & (2)      &   (3)       & (4)  & (5)  & (6) & (7) & (8) & (9)\\\hline
PKS 0048-09 &   L   &   0.634   &   4.50E-09    &   2.2 &   46.66   &       &   9.6 &   4.97    \\
0133+476    &   F   &   0.859   &   9.59E-09    &   2.34    &   47.30   &   7.09    &   20.7    &   6.79    \\
1ES 0212+735    &   F   &   2.367   &   1.03E-09    &   2.85    &   47.62   &   4.16    &   8.5 &   9.23    \\
PKS 0215+015    &   F   &   1.715   &   5.97E-09    &   2.18    &   47.91   &       &       &   5.61    \\
3C 66A  &   L   &   0.444   &   2.49E-08    &   1.93    &   47.13   &   1.99    &   2.6 &       \\
4C +28.07   &   F   &   1.213   &   3.66E-09    &   2.52    &   47.28   &   7.29    &   16.1    &       \\
PKS 0235+164    &   L   &   0.94    &   3.27E-08    &   2.14    &   47.98   &   16.32   &   24  &   20.74   \\
PKS 0336-01 &   F   &   0.852   &   1.18E-09    &   2.5 &   46.37   &   19.01   &   17.4    &   5.85    \\
PKS 0420-01 &   F   &   0.916   &   5.65E-09    &   2.42    &   47.14   &   11.72   &   19.9    &   7.49    \\
PKS 0422+00 &   L   &   0.31    &   1.04E-09    &   2.38    &   45.21   &   1.7 &       &   6.11    \\
\\
PKS 0521-36 &   F   &   0.057   &   2.85E-09    &   2.6 &   43.92   &       &       &   1.83    \\
PKS 0528+134    &   F   &   2.06    &   4.01E-09    &   2.64    &   47.98   &   14.22   &   31.2    &   19.84   \\
PKS 0605-08 &   F   &   0.872   &   1.87E-09    &   2.43    &   46.60   &   4.53    &   7.6 &   4.05    \\
S5 0716+714 &   L   &   0.3 &   1.31E-08    &   2.15    &   46.35   &       &   10.9    &       \\
PKS 0723-008    &   L   &   0.128   &   5.97E-10    &   2.3 &   44.11   &   2.5 &       &       \\
PKS 0735+17 &   L   &   0.424   &   4.42E-09    &   2.02    &   46.29   &   3.17    &   3.8 &       \\
PKS 0736+01 &   F   &   0.189   &   2.31E-09    &   2.63    &   44.98   &   3.08    &   8.6 &       \\
PKS 0754+100    &   L   &   0.266   &   1.98E-09    &   2.39    &   45.33   &   5.52    &   5.6 &   7.33    \\
PKS 0808+019    &   L   &   1.148   &   1.07E-09    &   2.45    &   46.68   &       &       &   5.39    \\
B3 0814+425 &   L   &   0.53    &   8.73E-09    &   2.15    &   46.78   &   5.84    &   4.6 &       \\
\\
B2 0827+243 &   F   &   0.94    &   1.30E-09    &   2.79    &   46.52   &   15.46   &   13.1    &       \\
PKS 0829+046    &   L   &   0.174   &   2.47E-09    &   2.5 &   44.96   &       &       &   3.8 \\
4C +71.07   &   F   &   2.172   &   1.24E-09    &   2.98    &   47.63   &   10.67   &   16.3    &       \\
OJ 287  &   L   &   0.306   &   2.75E-09    &   2.38    &   45.62   &   18.03   &   17  &   7.76    \\
B2 0954+25A &   F   &   0.708   &   7.01E-10    &   2.41    &   45.94   &   4.83    &   4.3 &       \\
S4 0954+55  &   F   &   0.896   &   1.05E-08    &   2.05    &   47.46   &   4.63    &       &       \\
S4 0954+65  &   L   &   0.368   &   5.43E-10    &   2.51    &   45.08   &   6.62    &   6.2 &   5.93    \\
PKS 1055+01 &   F   &   0.89    &   7.14E-09    &   2.29    &   47.23   &   7.78    &   12.2    &   7.49    \\
PKS 1127-14 &   F   &   1.184   &   2.42E-09    &   2.73    &   47.08   &       &       &   3.22    \\
4C +29.45   &   F   &   0.724   &   5.30E-09    &   2.37    &   46.85   &   9.42    &   28.5    &   9.63    \\
\\
B2 1215+30  &   L   &   0.13    &   6.66E-09    &   1.98    &   45.33   &       &       &   0.94    \\
1219+285    &   H   &   0.102   &   6.92E-09    &   2.06    &   45.08   &   1.56    &   1.2 &       \\
3C 273  &   F   &   0.158   &   9.55E-09    &   2.75    &   45.40   &   5.71    &   17  &   6.05    \\
3C 279  &   F   &   0.536   &   3.24E-08    &   2.32    &   47.31   &   16.77   &   24  &   4.16    \\
B2 1308+32  &   F   &   0.996   &   6.76E-09    &   2.3 &   47.33   &   11.38   &   15.4    &   11.58   \\
PKS 1335-127    &   F   &   0.539   &   2.14E-09    &   2.5 &   46.10   &       &       &   6.38    \\
PKS 1406-076    &   F   &   1.494   &   1.77E-09    &   2.42    &   47.21   &   8.26    &       &       \\
PKS 1502+106    &   F   &   1.839   &   6.70E-08    &   2.22    &   49.04   &   11.13   &   12  &       \\
PKS 1510-08 &   F   &   0.36    &   4.86E-08    &   2.41    &   47.03   &   13.18   &   16.7    &   7.64    \\
B2 1611+34  &   F   &   1.397   &   5.38E-10    &   2.29    &   46.62   &   5.04    &   13.7    &   3.36    \\
\\
B3 1633+382     &   F   &   1.814   &   6.78E-09    &   2.47    &   48.03   &   8.83    &   21.5    &   5.29    \\
PKS 1717+177    &   L   &   0.137   &   4.72E-09    &   2.01    &   45.20   &       &       &   1.94    \\
PKS 1725+044    &   F   &   0.296   &   1.27E-09    &   2.65    &   45.18   &   2.46    &   3.8 &       \\
PKS 1730-13 &   F   &   0.902   &   3.57E-09    &   2.34    &   46.93   &       &   10.7    &   11.84   \\
S5 1749+701 &   L   &   0.77    &   1.99E-09    &   2.05    &   46.57   &       &       &   3.75    \\
4C +09.57   &   L   &   0.322   &   6.45E-09    &   2.29    &   46.07   &   15.85   &   12  &       \\
8C 1803+784 &   L   &   0.68    &   3.04E-09    &   2.35    &   46.54   &   6.45    &   12.2    &   4.7 \\
3C 371  &   L   &   0.051   &   1.88E-09    &   2.6 &   43.64   &   1.8 &   1.1 &   1.05    \\
4C +56.27   &   L   &   0.664   &   2.67E-09    &   2.34    &   46.46   &       &   6.4 &   2.5 \\
PKS B1921-293   &   F   &   0.353   &   1.40E-09    &   2.4 &   45.46   &       &       &   9.51    \\
\\
8C 2007+777 &   L   &   0.342   &   1.43E-09    &   2.42    &   45.44   &   5.13    &   7.9 &   4.68    \\
4C -02.81   &   F   &   1.285   &   8.13E-10    &   2.31    &   46.70   &       &       &   7   \\
PKS 2145+06     &   F   &   0.99    &   7.48E-10    &   2.56    &   46.34   &   7.81    &   15.6    &   4.35    \\
PKS 2155-152    &   F   &   0.672   &   9.24E-10    &   2.51    &   45.98   &       &       &   2.31    \\
BL Lac  &   L   &   0.069   &   7.10E-09    &   2.38    &   44.56   &   3.91    &   7.3 &   2.77    \\
3C 446  &   F   &   1.404   &   2.15E-09    &   2.53    &   47.23   &   11.38   &   16  &   9.93    \\
PKS 2227-08 &   F   &   1.56    &   4.60E-09    &   2.65    &   47.69   &   12.42   &   15.9    &       \\
CTA 102 &   F   &   1.037   &   4.10E-09    &   2.56    &   47.14   &   14.23   &   15.6    &   8.02    \\
3C 454.3    &   F   &   0.859   &   4.62E-08    &   2.47    &   47.97   &   21.84   &   33.2    &   9.38    \\
\hline
\end{tabular}
 \,\,\,\,\,\,\,\,\,\,\,\,\,\,\,\,\,\,\,\,\,\,\,\,\,\,\,\,\,\,\,\,\,\,\,\,\,\,\,\,\,\,\,\,\,\,\,\,\,\,\,\,\,\,\,\,\,\,\,\,
 \,Note: F09: Fan et al.(2009); H09: Hovatta et al.(2009); L99: Lahteenimaki \& Valtaoja (1999)
\label{tab:samp}
\end{center}
\end{table*}

\subsection{Results}

For a source, the $\gamma$-ray luminosity can be calculated from
the detected photons. Here, the integral  luminosity is used in
our discussion since it is a more   robust measure of the
gamma-ray output (See Abdo et al. 2010c).

Let
 $${\rm{\frac{dN}{dE}} = N_{0} E^{-\alpha_{ph}}},$$
here $\alpha_{\rm{ph}}$ is the photon spectral index, and
$\rm{N_{0}}$ can be expressed as
 $$ \rm{N_{0} =  N_{(E_L\sim E_U)}({\frac{1}{E_L}-\frac{1}{E_U}}), if \alpha_{ph} =2, otherwise } $$
 $$ \rm{N_{0} = {\frac{N_{(E_{L}\sim E_{U})}(1-\alpha_{ph})}{(E_{U}^{1-\alpha_{ph}}-E_{L}^{1-\alpha_{ph}})}}, } $$
where $\rm{N_{(E_{L}\sim E_{U})}}$ is the integral photons in
units of $\rm photons \cdot cm^{-2}\cdot s^{-1}$ in the energy
range of $\rm{E_{L}}$ - $\rm{E_{U}}$. Therefore, the flux can be
obtained by $f = \rm \int^{E_U}_{E_L} E dN$, which can be
expressed as
  $$f = \rm{ N_{(E_L\sim E_U)}({\frac{1}{E_L}-\frac{1}{E_U}})ln{\frac{E_U}{E_L}}, if \alpha_{ph} = 2,
  otherwise}$$
  $$ f = \rm{N_{(E_L\sim E_U)}\frac{1-\alpha_{ph}}{2-\alpha_{ph}} \frac{(E_{U}^{2-\alpha_{ph}}-E_{L}^{2-\alpha_{ph}})}{(E_{U}^{1-\alpha_{ph}}-E_{L}^{1-\alpha_{ph}})}   }$$
in units of $\rm GeV \cdot cm^{-2}\cdot s^{-1}$. So, we can get
the $\gamma$-ray luminosity by
 $$L_{\gamma} = 4\pi d_L^2(1+z)^{(\alpha_{ph}-2)}f,$$
here $d_L$ is the luminosity distance, and can be expressed in the
form
$$d_L =
 \frac{c}{H_{0}}\int^{1+z}_{1}\frac{1}{\sqrt{\Omega_{M}x^{3}+1-\Omega_{M}}}
\rm dx$$
  from the $\Lambda-CDM$ model (Pedro \& Priyamvada, 2007)
with $\Omega_{\Lambda}\simeq0.7$, $\Omega_{M}\simeq0.3$ and
$\Omega_{K}\simeq0.0$, and $(1+z)^{(\alpha_{ph}-2)}$ represents a
K-correction.
 The calculated luminosity is listed in Col. 6
in Table 1 for 59 Fermi sources.  From the data listed in Table 1,
  we can get the average values for $\gamma$-ray luminosity as follows:\\
  $<logL_{\gamma}|^{FSRQs}>$ = $46.85\pm1.00$erg/s for the   36 FSRQs, and \\
  $ <log  L_{\gamma}|^{LBLs}> $  = $ 45.81 \pm 1.04 $ erg/s  for the
  22 LBLs.
  For HBL, there is only one source 1219+285, its $\gamma$-ray luminosity is $log\, L_{\gamma}   =  45.08 $
  erg/s.
  If it represents the average luminosity of HBLs, then
  the average values of $\rm{log L_{\gamma}}$ suggest a sequence  that
  $ log  L_{\gamma}|^{FSRQs} \,\,>\,\, log  L_{\gamma}|^{LBLs} \,\,>\,\,  log  L_{\gamma}|^{HBLs}.$

\begin{figure*}
\vbox to7.2in{\rule{0pt}{7.2in}}
\includegraphics{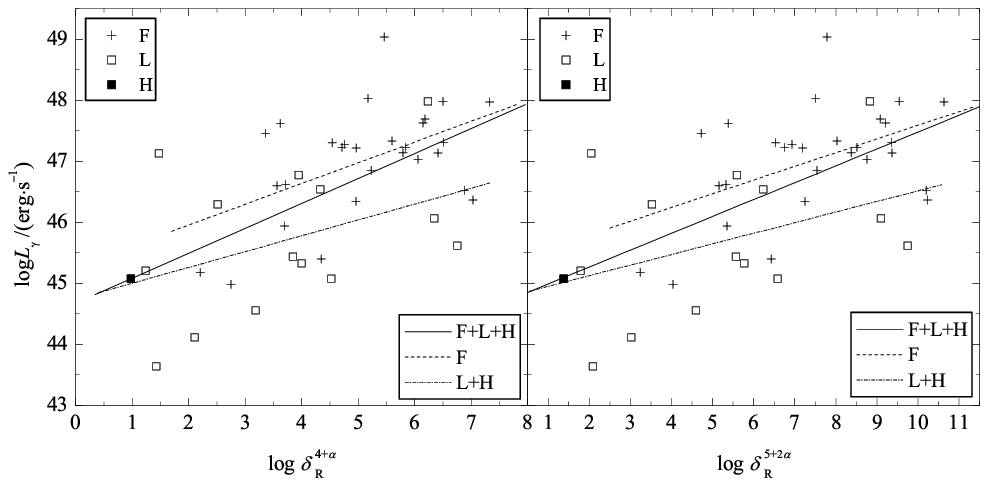} \caption{Plot of  the $\gamma$-ray
luminosity, log $\rm{L_{\nu}}$ (ergs/s) against  log
$\delta_R^{4+\alpha_{\gamma}}$ on the left panel and against log
$\delta_R^{5+2\alpha_{\gamma}}$ on the right panel for the sources
whose radio Doppler factors are from L\"{a}hteenm\"{a}ki \&
Valtaoja (1999).
 The plus  stands for FSRQs,
the open square  stands  for LBLs, and the filled squares   for
HBLs. The lines are for best fitting results. The solid line
stands for the whole sample (F+L+H), the dotted line for FSRQs
(F), the broken-line for BLs (H+L).}

\label{Fig1-PASJ}
\end{figure*}

For the beaming effect in the $\gamma$-ray emissions, we discussed
it using the $\gamma$-ray luminosity, $\rm{log L_{\gamma}}$ and
the radio Doppler factor,
  $\delta_R$ by discussing the linear
correlation between $\rm log L_{\gamma} - log\delta^{4+\alpha}
(or\,\, log\delta^{5+2\alpha})$. The Pearson's correlation
coefficient $r$ is expressed as (see Press 1994, Pavlidou et al.
2012)
 $$r = {\frac{\sum (x_i-\bar{x})(y_i-\bar{y})}{\sqrt{\sum(x_i-\bar{x})^2}\sqrt{\sum(y_i-\bar{y})^2}}}$$
here, $\bar{x}$ is the mean of the $x_i$'s, $\bar{y}$ is the mean
of the $y_i$'s, and ($x_i$,$y_i$) correspond to
(log$\delta_i^{4+\alpha}$ (or log$\delta^{5+2\alpha}$), log $\rm
L_{\gamma,i}$).
 Since the radio Doppler factors are from 3
different literatures, we considered the relationship between the
$\gamma$-ray luminosity, $\rm{log L_{\gamma}}$ and the radio
Doppler factor, $\delta_R$ for 3 samples separately, the results
are:

 $\rm log L_{\gamma}(erg/s) =
(0.41\pm0.09)log\delta_R^{4+\alpha_{\gamma}} + (44.67\pm 0.42)$
with a correlation coefficient $r = 0.597 $ and a chance
probability of $p<10^{-4}$, and
 $\rm log L_{\gamma}(erg/s) = (0.28\pm0.06)log\delta_R^{5+2\alpha_{\gamma}} +
(44.71\pm0.42)$ with a correlation coefficient $r = 0.590 $ and a
chance probability of $p<10^{-4}$ for the whole sample of 43
sources from L\"{a}hteenm\"{a}ki \& Valtaoja (1999).
 The corresponding plots are shown in Fig \ref{Fig1-PASJ}.

\begin{figure*}
\vbox to7.2in{\rule{0pt}{7.2in}}
\includegraphics{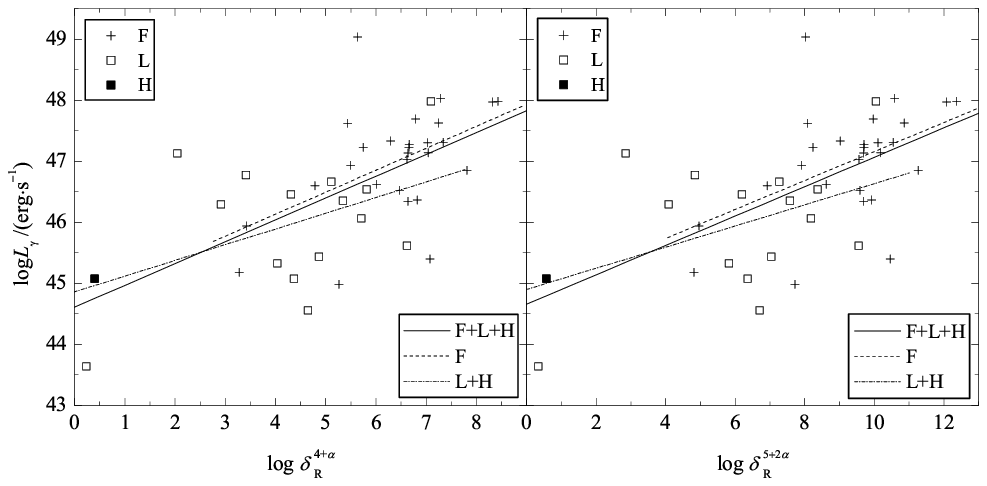} \caption{Plot of  the $\gamma$-ray
luminosity, log $\rm{L_{\nu}}$ (ergs/s) against  log
$\delta_R^{4+\alpha_{\gamma}}$ on the left panel and against log
$\delta_R^{5+2\alpha_{\gamma}}$ on the right panel for the sources
whose radio Doppler factors are from Hovatta et al. (2009). The
symbols and lines have the same meanings as in Fig. 1.}
\label{Fig2-PASJ}
\end{figure*}

 $\rm log L_{\gamma}(erg/s) =
(0.36 \pm0.07)log \delta_R^{4+\alpha_{\gamma}} + (44.61\pm0.42)$
with a correlation coefficient $r = 0.616 $ and a chance
probability of $p<10^{-4}$, and
 $\rm log L_{\gamma}(erg/s) =
 (0.24\pm0.05)log\delta_R^{5+2\alpha_{\gamma}} +
(44.66\pm0.42)$ with a correlation coefficient $r = 0.607 $ and a
chance probability of $p<10^{-4}$ for the whole sample of 43
sources from Hovatta et al. (2009).  The corresponding plots are
shown in Fig \ref{Fig2-PASJ}.

\begin{figure*}
\vbox to7.2in{\rule{0pt}{7.2in}}
\includegraphics{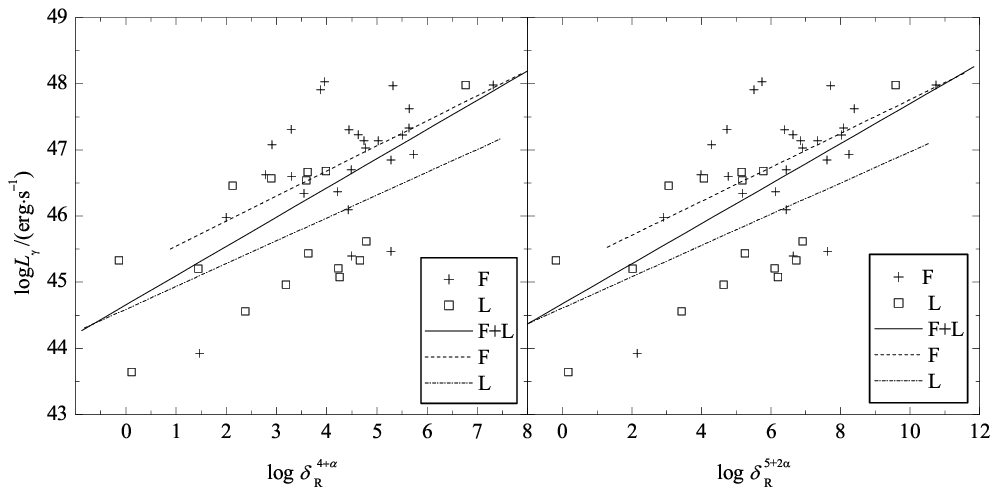} \caption{Plot of  the $\gamma$-ray
luminosity, log $\rm{L_{\nu}}$ (ergs/s) against  log
$\delta_R^{4+\alpha_{\gamma}}$ on the left panel and against log
$\delta_R^{5+2\alpha_{\gamma}}$ on the right panel for the sources
whose radio Doppler factors are from Fan et al. (2009). The
symbols and lines have the same meanings as in Fig. 1.}
\label{Fig3-PASJ}
\end{figure*}

$\rm log L_{\gamma}(erg/s) =
(0.44\pm0.09)log\delta_R^{4+\alpha_{\gamma}} + (44.65\pm0.37)$
with a correlation coefficient $r = 0.624 $ and a chance
probability of $p<10^{-4}$, and
 $\rm log L_{\gamma}(erg/s) =
 (0.30\pm0.06)log\delta_R^{5+2\alpha_{\gamma}} +
(44.67\pm0.37)$ with a correlation coefficient $r = 0.620 $ and a
chance probability of $p<10^{-4}$ for the whole sample of  42
sources from Fan et al. (2009).  The corresponding plots are shown
in Fig \ref{Fig3-PASJ}.

 For each sample, we also investigated
the correlation for the subclasses of FSRQs, and BLs, the
corresponding results are listed in Table 2. In the Table,
 column (1) gives the relationship,
 column (2) sample,  T for the whole sample, F for FSRQs, H+L for BLs,
 column (3) regression  constant $a$,
 column (4) 1 $\sigma$ uncertainty for constant $a$,
 column (5) slope $b$,
 column (6) 1 $\sigma$ uncertainty for slope $b$,
 column (7) correlation coefficient $r_{L\delta}$,
 column (8) number of sources, N,
 column (9) chance probability $p$,
 column (10) reference for the used radio Doppler factor, L99: L\"{a}hteenm\"{a}ki \& Valtaoja (1999);
    H09: Hovatta et al. (2009); and
    F09: Fan et al. (2009).

\begin{table*}
 \begin{center}
 Table~ 2. \hspace{20pt} Correlations between the $\gamma$-ray luminosity and the Radio Doppler Factor
  \vspace{6pt}
\begin{tabular}{lccccccccc}
\hline
 Relat. & Samp.  & a & $\Delta$a &  b & $\Delta$b & $r_{L\delta}$ & N & $p$
 & Ref for $\delta$
 \\
    (1)  & (2)      &   (3)       & (4)  & (5)  & (6) & (7) & (8) & (9) & (10)\\\hline
log$\delta^{4+\alpha}$-log L$_{\gamma}$ &   T   &   44.67   &   0.42    &   0.41    &   0.09    &   0.597   &   43  &   $<0.0001$   &   L99 \\
log$\delta^{4+\alpha}$-log L$_{\gamma}$ &   F   &   45.27   &   0.59    &   0.34    &   0.11    &   0.510   &   28  &   0.00556 &   L99 \\
log$\delta^{4+\alpha}$-log L$_{\gamma}$ &   H+L &   44.74   &   0.61    &   0.26    &   0.15    &   0.428   &   15  &   0.11153 &   L99 \\
log$\delta^{5+2\alpha}$-log L$_{\gamma}$    &   T   &   44.71   &   0.42    &   0.28    &   0.06    &   0.590   &   43  &   $<0.0001$   &   L99 \\
log$\delta^{5+2\alpha}$-log L$_{\gamma}$    &   F   &   45.34   &   0.59    &   0.22    &   0.08    &   0.495   &   28  &   0.00744 &   L99 \\
log$\delta^{5+2\alpha}$-log L$_{\gamma}$    &   H+L &   44.78   &   0.61    &   0.17    &   0.11    &   0.413   &   15  &   0.12631 &   L99 \\
    &       &       &       &       &       &       &       &       &       \\
log$\delta^{4+\alpha}$-log L$_{\gamma}$ &   T   &   44.61   &   0.42    &   0.36    &   0.07    &   0.616   &   43  &   $<0.0001$   &   H09 \\
log$\delta^{4+\alpha}$-log L$_{\gamma}$ &   F   &   44.69   &   0.83    &   0.36    &   0.13    &   0.489   &   27  &   0.00963 &   H09 \\
log$\delta^{4+\alpha}$-log L$_{\gamma}$ &   H+L &   44.86   &   0.58    &   0.26    &   0.13    &   0.477   &   16  &   0.0615  &   H09 \\
log$\delta^{5+2\alpha}$-log L$_{\gamma}$    &   T   &   44.66   &   0.42    &   0.24    &   0.05    &   0.607   &   43  &   $<0.0001$   &   H09 \\
log$\delta^{5+2\alpha}$-log L$_{\gamma}$    &   F   &   44.79   &   0.84    &   0.24    &   0.09    &   0.473   &   27  &   0.01271 &   H09 \\
log$\delta^{5+2\alpha}$-log L$_{\gamma}$    &   H+L &   44.90   &   0.59    &   0.17    &   0.09    &   0.461   &   16  &   0.07205 &   H09 \\
    &       &       &       &       &       &       &       &       &       \\
log$\delta^{4+\alpha}$-log L$_{\gamma}$ &   T   &   44.65   &   0.37    &   0.44    &   0.09    &   0.624   &   42  &   $<0.0001$   &   F09 \\
log$\delta^{4+\alpha}$-log L$_{\gamma}$ &   F   &   45.17   &   0.58    &   0.38    &   0.13    &   0.522   &   26  &   0.00622 &   F09 \\
log$\delta^{4+\alpha}$-log L$_{\gamma}$ &   H+L &   44.59   &   0.47    &   0.35    &   0.13    &   0.580   &   16  &   0.01854 &   F09 \\
log$\delta^{5+2\alpha}$-log L$_{\gamma}$    &   T   &   44.67   &   0.37    &   0.30    &   0.06    &   0.620   &   42  &   $<0.0001$   &   F09 \\
log$\delta^{5+2\alpha}$-log L$_{\gamma}$    &   F   &   45.20   &   0.58    &   0.26    &   0.09    &   0.516   &   26  &   0.00701 &   F09 \\
log$\delta^{5+2\alpha}$-log L$_{\gamma}$    &   H+L &   44.61   &   0.48    &   0.24    &   0.09    &   0.569   &   16  &   0.02155 &   F09 \\
\hline
\end{tabular}
 \,\,\,\,\,\,\,\,\,\,\,\,\,\,\,\,\,\,\,\,\,\,\,\,\,\,\,\,\,\,\,\,\,\,\,\,\,\,\,\,\,\,\,\,\,\,\,\,\,\,\,\,\,\,\,\,\,\,\,\,
 \,Note: log $L_{\gamma}$ = (a $\pm \Delta$a) + (b$\pm \Delta$b)log $\delta^q$, $q = 4+\alpha (\rm or\,\, 5+2\alpha)$
\label{tab:corr}
\end{center}
\end{table*}

\section{Discussion}

Blazars are a special subclass of AGNs showing extreme
observational properties, which are believed to be due to the
beaming effect. The beaming model was adopted to explain both the
particularly observational similarities and some observational
differences between BL lacertae objects and FSRQs.
  For the
two types of BL Lacertae objects (RBLs and XBLs),
  the beaming effect can explain some of the observational differences
  between them
  (see
 Fan et al.  1997;
 Fan \& Xie 1996;
 Georganopoulos \& Marscher 1999;
 Xie et al. 1991) although
the viewing angle alone can not explain all the  difference between
 RBLs and XBLs (Sambruna et al. 1996, Fossati et al. 1998).

The Fermi mission has detected a lot of blazars (Abdo et al.
2010a, Ackermann et al. 2011a), which shed new lights on the
emission mechanisms of blazars, particularly on the highly
energetic $\gamma$-ray emissions. There are many indirect
evidences to show that the $\gamma$-ray emissions are strongly
beamed (see
 Arshakian et al., 2010,
 Fan et al. 2008,
 Huang, et al., 1999,
 Kovalev et al.,  2009,
 Pushkarev et al., 2010, and
 Savolainen et al., 2010).

\begin{figure*}
\vbox to7.2in{\rule{0pt}{7.2in}}
\includegraphics{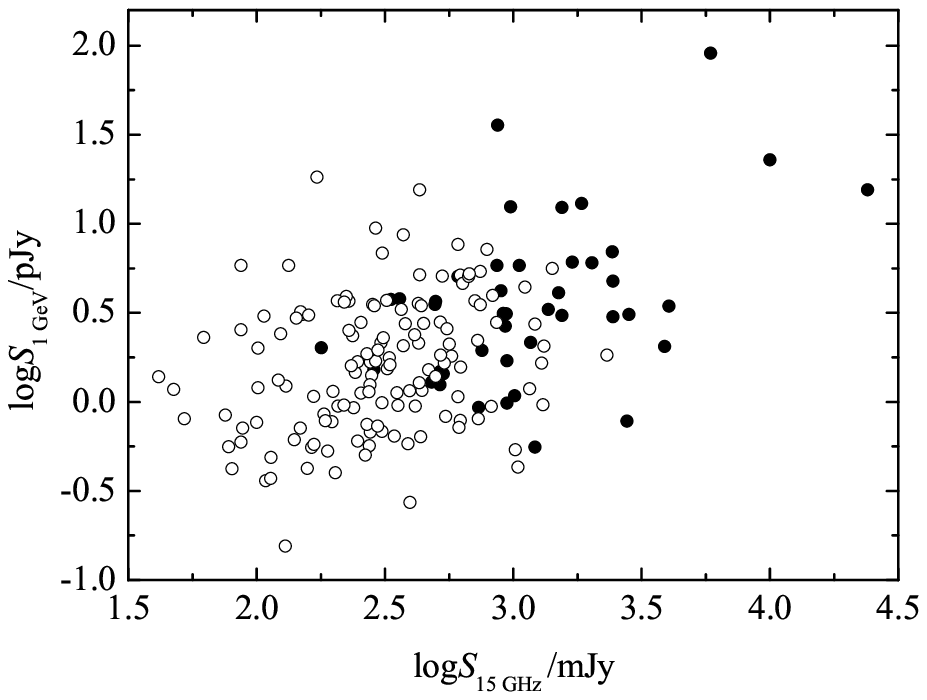} \caption{Plot of  the $\gamma$-ray
flux density, log $\rm{S_{1GeV}}$ (pJy) against the radio flux
density log $\rm S_{15GHz}$(mJy).  The 15GHz data are from the
paper by Ackermann et al.(2011b), the 43 black points stand for
the sources included in Table 1 while the open circles for the
sources that have 15GHz data  but not in our sample.}
\label{Fig4-PASJ}
\end{figure*}

 In the present paper, we chose the Fermi sources with
available radio Doppler factors for the discussion of the beaming
effect in the $\gamma$-ray region, and got a sample of 59 Fermi
sources. To show the representation of the sample, we put them in
a plot of the $\gamma$-ray flux density against the 15GHz radio
flux density(Fig. \ref{Fig4-PASJ}), the 15GHz radio flux densities
are from a paper by Ackermann et al. (2011b). Out of the 59
sources, 43 have corresponding 15 GHz radio data (namely 31
sources from both L\"{a}hteenm\"{a}ki \& Valtaoja 1999 and  Fan et
al. 2009, and 34 sources from Hovatta et al. 2009 have 15GHz radio
data). In the plot, the $\gamma$-ray flux density is calculated at
1GeV,  the 43 filled points stand for the sources included in
Table 1 of this paper  while the open circles for the rest sources
in the paper by Ackermann et al. (2011b). From the plot, it is
clear that the sources considered in the present paper show higher
$\gamma$-ray and 15 GHz radio flux densities than do the rest
sources. When the 8.4GHz radio data are used for a plot, similar
result can be obtained, namely the sources considered in the
present sample show higher $\gamma$-ray and 8 GHz radio flux
densities.

In a beaming model,  the $\gamma$-ray flux density
$S_{\nu}^{\rm{ob.}} $, is expected to be associated with the
Doppler factor, $\delta_{\nu}$ by  $S_{\nu}^{\rm{ob.}} \propto
\delta_{\nu}^{3+\alpha_{\nu}}$ in a synchrotron self-Compton (SSC)
model, or  $S_{\nu}^{\rm{ob.}} \propto
\delta_{\nu}^{4+2\alpha_{\nu}}$ in an external Compton (EC) model,
 here, $\alpha_{\nu}$ is the
spectral index ($f_{\nu} \propto \nu^{-\alpha_{\nu}}$)(see Dermer
1995).
 These indices are true for transient emission features,
     whereas in a steady jet, the indices are smaller by one.
   For the $\gamma$-ray luminosity, we can expect that
$L_{\gamma}^{\rm{ob.}} \propto \delta_{\nu}^{4+\alpha_{\gamma}}$
in a synchrotron self-Compton (SSC) model, and
$L_{\gamma}^{\rm{ob.}} \propto \delta_{\gamma}^{5+2\alpha_{\nu}}$
in an external Compton (EC) model. Therefore, for the
$\gamma$-ray sources, if we have a complete sample with available
$\gamma$-ray Doppler factor, $\delta_{\gamma}$, then we can use
the correlation between the $\gamma$-ray luminosity and the
$\gamma$-ray Doppler factor, $\delta_{\gamma}$, to check  the
emission mechanism for the $\gamma$-rays. Unfortunately, we do not
have $\delta_{\gamma}$ for the sources.  If the $\gamma$-ray
Doppler factor, $\delta_{\gamma}$, is the same as the radio
Doppler factors, $\delta_R$, then we can use the radio Doppler
factor in our consideration. In the present paper, we compiled the
radio Doppler factors from 3 papers(see L\"{a}hteenm\"{a}ki \&
Valtaoja 1999; Hovatta et al. 2009; Fan et al. 2009) for the
$\gamma$-ray sources(Abdo et al. 2010a) and got 3 corresponding
samples. For each sample, we made linear regression fitting for
the whole sample, and the sub-samples for FSRQs and BL Lacs
(LBLs+HBLs) respectively.  Significant correlations are
obtained for all the relations(see Fig.s 1, 2, and 3, and Table
2).  However, Table 2 shows that all the slopes are much below the
expected value, 1.0. The reasons are probably that 1) the present
sample  is too small; 2) the $\gamma$-ray Doppler factors are not
the same as the radio Doppler factors;  3) the $\gamma$-ray
emissions and the radio emissions used for the radio Doppler
factors are not simultaneous; or 4) the correlations are from the
effect of a common redshift as discussed below.

  If the correlation is an apparent one caused by the
redshift, which is correlated with the luminosity and the Doppler
factor, $\delta$, then it is important for us to remove
  the effect of a  common redshift.
  To do so, we can use the method (Padovani 1992) to deal
with the relevant data. If $r_{ij}$ is the correlation coefficient
between x$_i$ and x$_j$, in the case of three variables the
correlation coefficient between two of them, removing the effect
of the third one is (see also Fan et al. 1996)
$$ r_{12,3}={\frac{r_{12}-r_{13}r_{23}}{\sqrt{(1-r_{12}^2)(1-r_{22}^2)}}}.$$
  We  adopted this formulae  to the analysis, and listed the results
in Table \ref{tab:corr-z}, in which,
 Col. (1) gives the correlations,
 Col. (2) the sample,
 Col. (3) correlation coefficient for luminosity and the radio
 Doppler factor, the values are the same as that in Col. (7) of
 Table 2,
 Col. (4) correlation coefficient for redshift and the radio
 Doppler factor ($\rm log z\, - \,log \delta^{4+\alpha}$, or $\rm log z\, - \,log
 \delta^{5+2\alpha}$),
 Col. (5) correlation coefficient for redshift and the
 $\gamma$-ray luminosity, $\rm log L_{\gamma}\, - \, log z$),
 Col. (6) correlation coefficient removing the effect of a common
 redshift,
 Col. (7) number of sources, N,
 Col. (8) chance probability, $p$,
 Col. (9) the existence of the correlation,
   'No' means that the  correlation, after removing the effect of a common redshift, does not exist any more;
   'Mar'. means there is a marginal correlation  after removing the effect of a common
  redshift;
   'Yes' means that there is still a correlation after removing the effect of a common
  redshift,
 Col. (10) the reference for the radio Doppler factors.

From Table  \ref{tab:corr-z}, for the whole sample, we have
$r_{L\delta,z}$ = 0.245 (L99), 0.211 (H99), and 0.243 (F09) for
the case of log$\delta^{4+\alpha}$-log L$_{\gamma}$, and
 $r_{L\delta,z}$ = 0.222 (L99), 0.187 (H99), and 0.232 (F09) for the case of
log$\delta^{5+2\alpha}$-log L$_{\gamma}$. The chance probability
is greater than 10\%, suggesting that there is no more
correlation.

For FSRQs, we have
 $r_{L\delta,z}$ = 0.422 (L99), 0.480 (H99), and 0.363 (F09) for the case of
log$\delta^{4+\alpha}$-log L$_{\gamma}$, and
 $r_{L\delta,z}$ = 0.395 (L99), 0.451 (H99), and 0.353 (F09) for the case of
log$\delta^{5+2\alpha}$-log L$_{\gamma}$. The  chance probability
is less than 5\% for the L99 and H09 samples, suggesting that the
correlation exists for FSRQs. The   chance probability are 6.6\%
and 7.4\% for F09 sample, implying a marginal correlation. We can
say that there is a correlation between the $\gamma$-ray
luminosity and the radio Doppler factor ($\delta^{4+\alpha}$ or
$\delta^{5+2\alpha}$). This result implies that the $\gamma$-ray
emissions are really correlated with the Doppler factors, and that
the $\gamma$-ray Doppler factors are associated with the radio
Doppler factor in FSRQs. We also found, for each sample, that
there is no much difference between the correlation coefficient
$r_{L\delta,z}$ for $\rm log\delta^{3+\alpha}-log L_{\gamma}$ and
that for $\rm log \delta^{4+2\alpha} - log L_{\gamma}$. In
addition, there is no much difference in the correlation
coefficient $r_{L\delta,z}$ for the 3 samples. So, based on the
analysis, it is difficult for us to tell one emission mechanism
(SSC or EC) from another for FSRQs. The reasons are perhaps 1) the
present samples are small and not complete and 2) the $\gamma$-ray
Doppler factors are not the same as the radio Doppler factor.

\begin{table*}
 \begin{center}
 Table~ 3. \hspace{20pt} Correlations removing the effect of a  common redshift
  \vspace{6pt}
\begin{tabular}{lccccccccc}
\hline
 Relat. & Samp.  & $r_{L\delta}$ & $r_{z\delta}$ & $r_{zL}$  & $r_{L\delta,z}$ &  N & $p$& Corr& Ref for $\delta$
 \\
    (1)  & (2)      &   (3)       & (4)  & (5)  & (6) & (7) & (8)&(9) &(10)\\\hline
log$\delta^{4+\alpha}$-log L$_{\gamma}$ &   T   &   0.597   &   0.570   &   0.879   &   0.245   &   43  &   0.116   &   No   &   L99 \\
log$\delta^{4+\alpha}$-log L$_{\gamma}$ &   F   &   0.510   &   0.333   &   0.774   &   0.422   &   28  &   0.026   &   Yes   &   L99 \\
log$\delta^{4+\alpha}$-log L$_{\gamma}$ &   H+L &   0.428   &   0.511   &   0.877   &   -0.048  &   15  &   0.846   &   No   &   L99 \\
log$\delta^{5+2\alpha}$-log L$_{\gamma}$    &   T   &   0.590   &   0.572   &   0.879   &   0.222   &   43  &   0.149   &   No   &   L99 \\
log$\delta^{5+2\alpha}$-log L$_{\gamma}$    &   F   &   0.495   &   0.335   &   0.774   &   0.395   &   28  &   0.039   &   Yes   &   L99 \\
log$\delta^{5+2\alpha}$-log L$_{\gamma}$    &   H+L &   0.413   &   0.502   &   0.877   &   -0.065  &   15  &   0.816   &   No   &   L99 \\
    &       &       &       &       &   ¡¡  &       &       &       &       \\
log$\delta^{4+\alpha}$-log L$_{\gamma}$ &   T   &   0.616   &   0.614   &   0.868   &   0.211   &   43  &   0.168   &   No   &   H09 \\
log$\delta^{4+\alpha}$-log L$_{\gamma}$ &   F   &   0.489   &   0.254   &   0.780   &   0.480   &   27  &   0.013   &   Yes   &   H09 \\
log$\delta^{4+\alpha}$-log L$_{\gamma}$ &   H+L &   0.477   &   0.586   &   0.882   &   -0.105  &   16  &   0.689   &   No   &   H09 \\
log$\delta^{5+2\alpha}$-log L$_{\gamma}$    &   T   &   0.607   &   0.616   &   0.868   &   0.187   &   43  &   0.213   &   No   &   H09 \\
log$\delta^{5+2\alpha}$-log L$_{\gamma}$    &   F   &   0.473   &   0.257   &   0.780   &   0.451   &   27  &   0.019   &   Yes   &   H09 \\
log$\delta^{5+2\alpha}$-log L$_{\gamma}$    &   H+L &   0.461   &   0.578   &   0.882   &   -0.127  &   16  &   0.603   &   No   &   H09 \\
    &       &       &       &       &   ¡¡  &       &       &       &       \\
log$\delta^{4+\alpha}$-log L$_{\gamma}$ &   T   &   0.624   &   0.600   &   0.896   &   0.243   &   42  &   0.120   &   No   &   F09 \\
log$\delta^{4+\alpha}$-log L$_{\gamma}$ &   F   &   0.522   &   0.407   &   0.835   &   0.363   &   26  &   0.066   &   Mar.    &   F09 \\
log$\delta^{4+\alpha}$-log L$_{\gamma}$ &   H+L &   0.580   &   0.641   &   0.899   &   0.012   &   16  &   0.968   &   No   &   F09 \\
log$\delta^{5+2\alpha}$-log L$_{\gamma}$    &   T   &   0.620   &   0.600   &   0.896   &   0.232   &   42  &   0.136   &   No   &   F09 \\
log$\delta^{5+2\alpha}$-log L$_{\gamma}$    &   F   &   0.516   &   0.405   &   0.835   &   0.353   &   26  &   0.074   &   Mar.    &   F09 \\
log$\delta^{5+2\alpha}$-log L$_{\gamma}$    &   H+L &   0.569   &   0.637   &   0.899   &   -0.011  &   16  &   0.968   &   No   &   F09 \\
\hline
\end{tabular}
\label{tab:corr-z}
\end{center}
\end{table*}

For Bls, however, there is no more correlation between the
luminosity and the Doppler factor if
  the effect of a  common
redshift is considered. The correlation coefficients are,
$r_{L\delta,z}$ = -0.048 (L99), -0.105 (H09), and 0.012(F09) for
the case of  $\rm log \delta^{4+\alpha} - log L_{\gamma}$,  and
$r_{L\delta,z}$ = -0.065 (L99), -0.127 (H09), and -0.011(F09) for
the case of $\rm log \delta^{5+2\alpha} - log L_{\gamma}$.
  The apparent correlation between the luminosity and the Doppler factor
is from   the effect of a  common redshift.
   Does that mean the $\gamma$-ray Doppler factors in BLs is quite
different from the radio Doppler factor? From above analysis, it
suggests that the $\gamma$-ray emission mechanism  in FSRQs is
different from that in BLs or the dependence of luminosity on
Doppler factor in FSRQs is different from that in BLs. We also
noticed that the BL sub-samples consist of only 15 or 16 objects,
the samples are too small.
  It is hard to draw a conclusion about the emission process
differences  between FSRQs and BL Lacs based on a sample of BL Lacs
that has only 15 or 16 sources.
  A complete sample with available
$\gamma$-ray Doppler factor should be obtained for the
investigation.

 When we  considered the correlation between the $\gamma$-ray
luminosity and the radio Doppler factor, $\rm log \delta - log
L_{\gamma}$, we have that there are correlations between them for
the whole sample and the sub-samples for each of the 3 samples.
The correlation analysis results are shown in Table 4. We can see
clearly that there is still a correlation between the $\gamma$-ray
luminosity and the radio Doppler factor for FSRQs even the effect
of a common redshift is removed,
 however the partial correlation analysis does not show a correlation for the small BL Lac
sample.

\begin{table*}
 \begin{center}
 Table~ 4. \hspace{20pt} Correlations for log$\delta$-log L$_{\gamma}$
  \vspace{6pt}
\begin{tabular}{lcccccccccccc}
\hline
 Relat. &
 Samp.  &
  a     &
 $\Delta$a &
   b       &
 $\Delta$b &
 $r_{L\delta}$ &
  N &
  $p$ &
  $r_{L\delta,z}$ &
   $p_z$ &
   Corr &
   Ref
 \\
    (1)  & (2)      &   (3)       & (4)  & (5)  & (6) & (7) & (8)&(9) &(10)&(11)&(12)& (13) \\\hline
log$\delta$-log L$_{\gamma}$    &   T   &   44.60   &   0.42    &   2.32    &   0.47    &   0.611   &   43  &   $<0.0001$   &   0.269   &   0.081  & Mar. &   L99 \\
log$\delta$-log L$_{\gamma}$    &   F   &   45.13   &   0.59    &   2.03    &   0.61    &   0.545   &   28  &   0.00271 &   0.386   &   0.049   &  Yes &  L99 \\
log$\delta$-log L$_{\gamma}$    &   H+L &   44.66   &   0.59    &   1.50    &   0.79    &   0.466   &   15  &   0.07999 &   0.069   &   0.783   & No &   L99 \\
    &       &       &       &       &       &       &       &       &   ¡¡  &       &       \\
log$\delta$-log L$_{\gamma}$    &   T   &   44.50   &   0.42    &   2.05    &   0.39    &   0.633   &   43  &   $<0.0001$   &   0.274   &   0.072   & Mar. &  H09 \\
log$\delta$-log L$_{\gamma}$    &   F   &   44.54   &   0.81    &   2.11    &   0.69    &   0.522   &   27  &   0.00523 &   0.548   &   0.004   & Yes &  H09 \\
log$\delta$-log L$_{\gamma}$    &   H+L &   44.76   &   0.57    &   1.48    &   0.65    &   0.517   &   16  &   0.04033 &   -0.048  &   0.847   & No & H09 \\
    &       &       &       &       &       &       &       &       &       &       &       \\
log$\delta$-log L$_{\gamma}$    &   T   &   44.63   &   0.37    &   2.44    &   0.47    &   0.632   &   42  &   $<0.0001$   &   0.301   &   0.048   & Yes&   F09 \\
log$\delta$-log L$_{\gamma}$    &   F   &   45.10   &   0.58    &   2.15    &   0.69    &   0.535   &   26  &   0.00482 &   0.491   &   0.009   & Yes &  F09 \\
log$\delta$-log L$_{\gamma}$    &   H+L &   44.55   &   0.46    &   1.91    &   0.67    &   0.607   &   16  &   0.01262 &   -0.004  &   0.988   & No&  F09 \\
\hline
\end{tabular}
 \,\,\,\,\,\,\,\,\,\,\,\,\,\,\,\,\,\,\,\,\,\,\,\,\,\,\,\,\,\,\,\,\,\,\,\,\,\,\,\,\,\,\,\,\,\,\,\,\,\,\,\,\,\,\,\,\,\,\,\,
 \,Note: log $L_{\gamma}$ = (a $\pm \Delta$a) + (b$\pm \Delta$b)log $\delta$
\label{tab:del-L}
\end{center}
\end{table*}

 In the present paper, we compiled 3 samples of Fermi loud blazars with
available radio Doppler factors. For each sample, we investigated
the correlation between the $\gamma$-ray luminosity, $\rm log
L_{\gamma}$, and the radio Doppler factor ($\rm log
\delta^{4+\alpha}$ or $\rm log \delta^{5+2\alpha}$) for the whole
sample, FSRQs and BLs respectively. Following conclusions can be
obtained. \\
 1. There are apparent correlations between the the $\gamma$-ray luminosity,
$\rm log L_{\gamma}$, and the radio Doppler factor ($\rm log
\delta^{4+\alpha}$ or $\rm log \delta^{5+2\alpha}$) for the whole
sample, FSRQs and BLs.\\
 2. When redshift effect  is excluded, there are still
  correlations between  the $\gamma$-ray luminosity,
$\rm log L_{\gamma}$, and the radio Doppler factor ($\rm log
\delta^{4+\alpha}$ or $\rm log \delta^{5+2\alpha}$) for  FSRQs
suggesting that there is a real correlation between the
$\gamma$-ray luminosity and the Doppler factor. However,
  the partial correlation analysis does not show a correlation
 for the small BL Lac sample.\\
 3. The $\gamma$-ray emission mechanism in FSRQs is perhaps
 different from that in BLs. Or there is a different dependence of
 the $\gamma$-ray emission on  the radio Doppler factor between FSRQs and BLs.

\section{Acknowledgements}
 The authors thank the anonymous referee for the constructive
suggestions and comments.  This work is partially supported by the
National 973 project (2007CB815405),  the National Natural Science
Foundation of China (NSFC 10633010, NSFC 11173009), and the Bureau
of Education  of Guangzhou Municipality(No.11 Sui-Jiao-Ke[2009]),
Guangdong Province Universities and Colleges Pearl River Scholar
Funded Scheme(GDUPS)(2009), Yangcheng Scholar Funded
Scheme(10A027S).

\end{document}